\documentclass[twocolumn,showpacs,preprintnumbers,amsmath,amssymb]{revtex4-1}
\usepackage{graphicx}
\usepackage{bm}
\usepackage{float}

\begin{document}

\title{Half-metallic Antiferromagnet BaCrFeAs$_{2}$}

\author{Shu-Jun Hu and Xiao Hu}
\affiliation{ World Premier International Center for Materials Nanoarchitectonics (MANA)\\
National Institute for Materials Science, Tsukuba 305-0044, Japan}

\date{\today}

\begin{abstract}
First-principles calculations and a tight-binding analysis predict
that the iron-pnictide BaCrFeAs$_{2}$ is a promising candidate for
half-metallic material with fully-compensated magnetization. The
transition-metal ions Cr and Fe prefer the three-dimensional
intervening lattice, which yields the antiferromagnetic
order of spin orientations. Due to the difference between Cr and Fe in the
electronegativity, a band gap is opened at the Fermi level in the spin channel
in which Fe provides the majority carriers. The selective hybridization between
3d orbitals of Cr and As:4p states due to the peculiar lattice
structure of the iron-pnictide is shown to be crucial for the novel properties.
\end{abstract}

\maketitle

\section{Introduction}
As one of the elements well explored from the ancient time, iron (Fe) had not
been a central player in the modern electronic technologies. However,
the recent discovery of superconductivity in iron-pnictide materials
\cite{Hosono} certainly has opened a new chapter of study on Fe. The
parent iron-pnictide materials are metallic and antiforromagentic (AFM),
and superconductivity is realized by carrier doping \cite{Hosono} or
application of pressure \cite{Pressure}.
So far the main focus on the iron-pnictide materials has been
superconductivity, and the AFM order is to be suppressed as a competitor. However,
it has been revealed during the investigation of the mechanism
of superconductivity that the AFM order of Fe ions itself is also
of interest which is established by the peculiar lattice structure
of iron-pnictides via the ferro-orbital ordering \cite{orbital-order}.
It will be fantastic if one can successfully explore the possibility of
iron-pnictides as a platform for novel spin-dependent transport properties
in which the magnetic order plays a direct role.

Of many possibilities, we focus here on a novel class of materials called
half-metallic antiferromagnet (HMAFM), which is metallic in one spin channel and
exhibits a gap at the Fermi level, thus insulating, in the other spin channel
while showing zero total magnetization in a unit cell \cite{deGroot-1st}.
HMAFM can generate spin-polarized current without perturbing the
surrounding elements magnetically in a device, and thus is particularly useful in
spintronics applications. HMAFM can also be the parent material for
single spin superconductivity with triplet Cooper pairs due to the finite density
of states in only one spin channel \cite{Single-spin}.
Since the first proposal \cite{deGroot-1st}, the possibility of HMAFM
has been investigated primarily in Heulser alloys \cite{heusler-2, heusler-3}
and double perovskites \cite{dp-1, dp-2, dp-3, dp-4, dp-5, dp-6, dp-7}.
The transition-metal (TM) chalcogenide superlattice \cite{monolayersl},
thiospinels \cite{thiospinel} and some disordered systems, such as
diluted magnetic semiconductors \cite{dms-1, dms-2}, vacancy-induced TM
oxides \cite{vacancy-tm-oxide} and hole-doped cuprates \cite{Pero-cuprate}
are also predicted to be the possible condicates.
Up to date, it is still lacking of a definite confirmation of such novel material.

It is intriguing to notice that the iron-pnictide materials can be
good candidates for HMAFM. The ground state of these materials is
AFM and poorly metallic. In order to achieve the HMAFM, one only needs
to open a gap in one spin channel while keeping
the AFM order with compensated magnetization. In the
simplest picture, a Fe atom has six 3d electrons and
shows an effective spin moment of 4 $\mu_{\textnormal B}$ due to the
intra-atomic Hund's coupling. In order to achieve HMAFM, one can
choose the Chromium (Cr) atom to substitute half of the Fe atoms
since Cr possesses four 3d electrons. It is expected
that the Cr atom will not change the AFM order of compensated
magnetization of the parent material, while modify the band
structure due to the different atomic number from Fe.

\begin{figure}[b]
\includegraphics[width=8.cm]{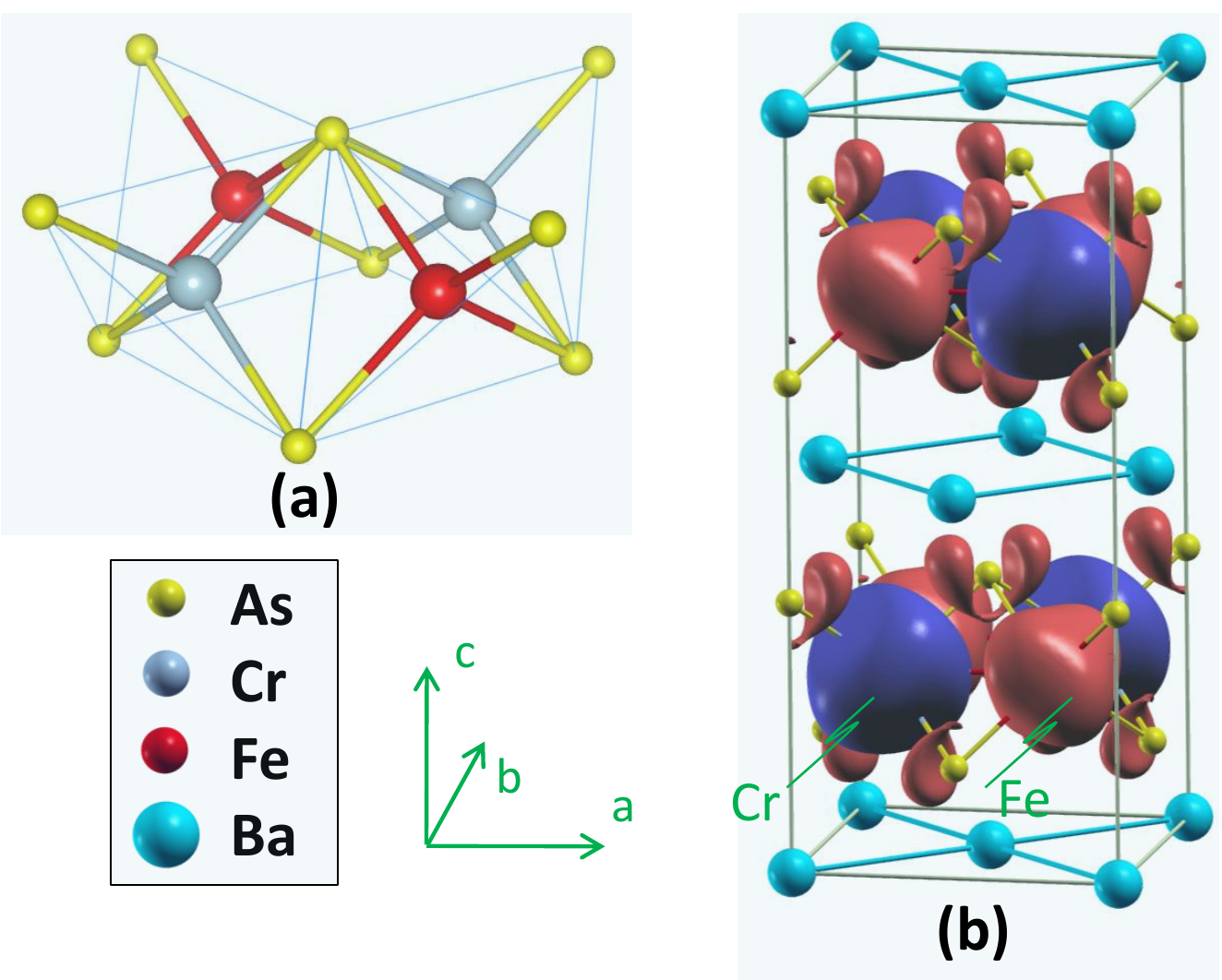}
\label{structure}
\caption{ (a) Crystal structure of the TM-As
layer of BaCrFeAs$_{2}$; (b) Spin-density of the AFM state
(blue/red surface: spin-up/spin-down).
Isovalue=0.025 e/\AA$^3$.}
\end{figure}

In the present work, we focus on a typical iron-pnictide BaFe$_{2}$As$_{2}$ \cite{first-122},
which belongs to the so-called '122' family. Among the well-studied iron-pnictides, the '122' family has
the shortest As-As bond along the c axis, which enhances the
interlayer coupling.
From first-principles calculations
on the electronic structure of BaCrFeAs$_{2}$, it is revealed that
(i) Cr and Fe ions prefer a three-dimensional (3D) intervening lattice; (ii) the magnetic moments
of Cr and Fe ions are aligned antiferromagnetically, and with small but non-negligible
contributions from As atoms parallel to that from Fe ions, the total
spin magnetization in a unit cell is zero; (iii) a gap is opened at
the Fermi level ($E_{\textrm F}$) in the spin channel where
Fe electrons contribute as majority carriers by the hybridization
between As:4p states and minority Cr:3d states due to the difference
in the electronegativity between Fe and Cr
as well as the peculiar crystal structure of iron-pnictide.
Based on these results, we predict BaCrFeAs$_{2}$ as a possible HMAFM.

This paper is organized as follows. In the next section the computational details of calculations
are briefly described. In the first part of section \uppercase\expandafter{\romannumeral3}, we focus on
the electronic structure of BaCrFeAs$_{2}$; then a tight-binding (TB) picture is presented. Finally the summary of
the paper is given in section \uppercase\expandafter{\romannumeral4}.

\section{Computational Details}
First-principles calculations were performed by VASP package
\cite{vasp-cite-1,vasp-cite-2} within the scheme of PAW method
\cite{PAW-method} and generalized gradient approximation \cite{PBE}
for the exchange-correlation functional
\footnote{It is widely accepted that the electronic structure of iron-pnictides
(particularly the Fermi surface) predicted by the local density approximation (LDA) or GGA
is in good agreement with the ARPES experiments.
In the current study we focus on the half-metallic bands in the vicinity of the
Fermi level. It is expected that GGA can predict the accurate band structures
of BaCrFeAs$_{2}$ and therefore the strong correlation effect (Hubbard U) is unnecessary}.
8$\times$8$\times$4 Monckhorst-Pack special points and 500 eV are
used for $k$ sampling and plane-wave basis set. Starting from the
experimental lattice parameters of BaFe$_{2}$As$_{2}$ in tetragonal
structure \cite{neutron122}, both the ion coordinations and lattice
parameters are fully relaxed until the force and stress are less
than 0.01 eV/\AA$ $ and 1 kBar. Calculations were also
performed by the Quantum-Espresso package \cite{PWscf}, and a good
agreement is achieved.

The on-site energies and hopping integrals of the TB model are
calculated by the maximally localized Wannier functions (MLWF)
method as implemented in the Wannier90 package \cite{Wannier90}
jointly with Quantum-Espresso \cite{PWscf}.
The Bloch wavefunction of BaCrFeAs$_{2}$ is first calculated in density functional formalism.
By specifying the 3d orbitals of TM ions and 4p orbitals of As ions as the initial guess,
the periodic wavefunctions are transformed to the MLWF representation in real space.
Meanwhile, the on-site energies and hopping integrals can be obtained.
It is noticed that the TB model including only the nearest neighboring hoppings can
reproduce well the DFT band structures.

\section{Results and Discussion}

\subsection{First-principles results}

\begin{figure}[t]
\includegraphics[width=8.cm]{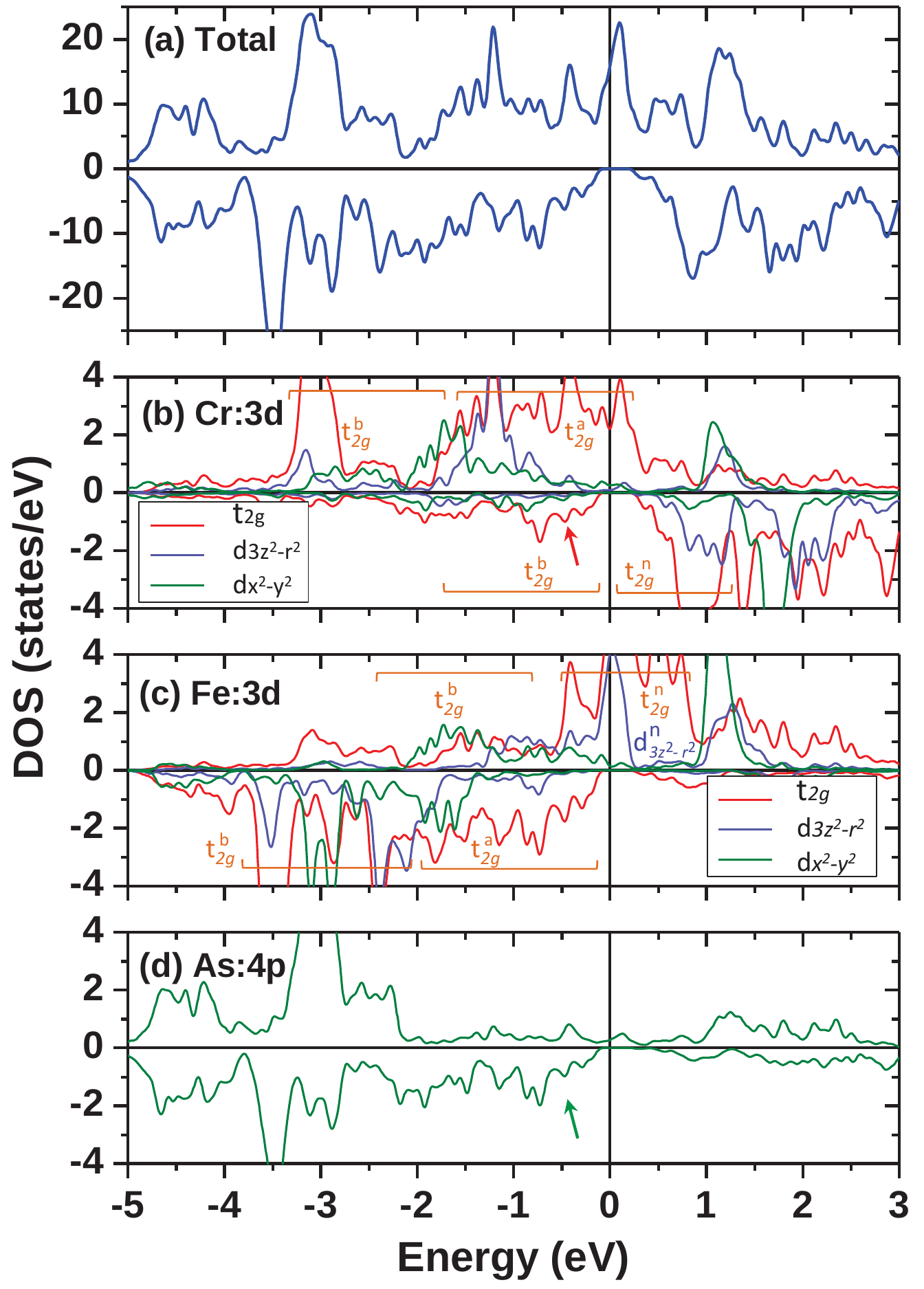}
\label{dos}
\caption{(a) Total density of states (DOS) of
half-metallic antiferromagnet BaCrFeAs$_{2}$. (b) and (c) Projected DOS of 3d states for
Cr and Fe, respectively. The Fermi energy $E_{\textrm F}$ is set to energy zero.
The bonding, anti-bonding and non-bonding states of t$_{2g}$ are labeled as t$_{2g}^{\textrm b}$,
t$_{2g}^{\textrm a}$ and t$_{2g}^{\textrm n}$, respectively. (d) Projected DOS of As:4p states.}
\end{figure}

In BaFe$_{2}$As$_{2}$, Fe ions tetrahedrally coordinated by As atoms
form Fe-As layers, which are sandwiched by
Ba layers. We concentrate on the situation that half of the Fe atoms are
substituted by Cr atoms in a $\sqrt{2}\times\sqrt{2}\times1$ cell.
The possible atomic configurations which have been studied include: (1)
cross-checkerboard type as shown in Figure 1b, (2) checkerboard type with
same positions of Cr and Fe ions along c axis, (3) alternating Cr and Fe
layers, and (4) alternating Cr and Fe stripes along a or b axis.
We calculate the total energies of them with all the possible spin orderings of
the TM ions. It is revealed that the ground state of BaCrFeAs$_{2}$ is characterized by the
intervening Fe and Cr ions and 3D AFM spin order, as shown in
Figure 1b. The resultant lattice parameters of a=5.725 \AA$ $ and c=13.290 \AA$ $
are in good agreement with the experiments \cite{BaCrFeAs2-exp}.

The magnetization is mainly contributed by Fe (-2.62 $\mu
_{\textrm B}$) and Cr (2.75 $\mu _{\textrm B}$) ions.
In addition, each As atom carries a small magnetic moment
(-0.08 $\mu _{\textrm B}$) parallel to that of Fe ions
due to the hybridization between As:4p and TM:3d electrons. The Ba layers act as charge reservoir,
and thus the magnetic moment of Ba$^{2+}$ ions is negligible.
According to the calculation, the magnetic moments of the Fe, Cr and
As atoms compensate completely, resulting in the total
magnetization of 0.0 $\mu _{\textrm B}$ per unit cell.

The total density of states (DOS) of BaCrFeAs$_{2}$ is depicted in
Figure 2a. While the spin-up channel exhibits an evident
metallic character similar with the parent material \cite{Singh-122band},
a gap of $\sim$ 0.3 eV emerges at $E_{\textrm F}$ in the spin-down
channel. Combining the zero magnetization and half-metallic band
structure, BaCrFeAs$_{2}$ is expected to be a HMAFM.

Let us look into the detailed electronic band structure of the new
material BaCrFeAs$_{2}$.
In the spin-up channel, Cr nominally has
four 3d electrons, and thus its majority 3d band is not fully occupied,
as illustrated by Figure 2b.
The hybridization between As:4p and Cr:t$_{2g}$ states forms the bonding
states (t$_{2g}^{\textrm b}$) deep in the valence band, and partially occupied
anti-bonding states (t$_{2g}^{\textrm a}$) around $E_{\textrm F}$.
Although there exist Fe:3d-As:4p as well as
Fe:3d$_{x^{2}-y^{2}}$-Cr:3d$_{x^{2}-y^{2}}$ hybridizations, most of
the Fe:3d minority states remain non-bonding and locate around $E_{\textrm F}$
(t$_{2g}^{\textrm n}$ and d$_{3z^{2}-r^{2}}^{\textrm n}$
in Figure 2c). BaCrFeAs$_{2}$ is therefore
metallic in the spin-up channel.

\begin{figure}[b]
\includegraphics[width=8cm]{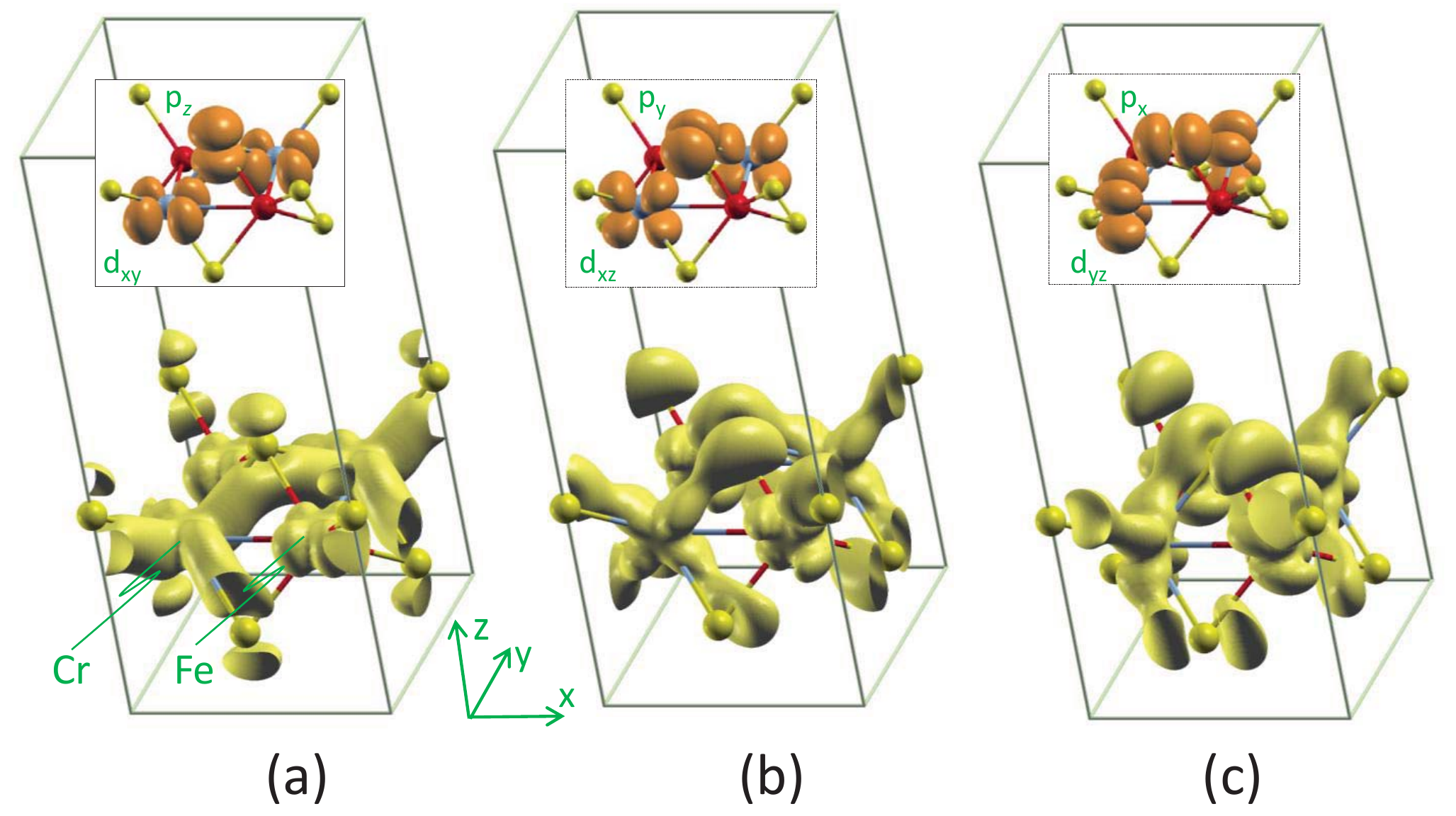}
\label{bdcd}
\caption{ Band-decomposed charge density for the
(a) Cr:3d$_{xy}$ bonding via As:4p$_z$, (b) Cr:3d$_{xz}$ bonding via
As:4p$_y$ and (c) Cr:3d$_{yz}$ bonding via As:4p$_x$ at the $\Gamma$ point
of the spin-down valence band maximum associated with a $\sqrt{2}\times\sqrt{2}\times1$ cell.
Only one TM-As layer is shown.
Isovalue=0.004 e/\AA$^3 $ for (a) and 0.008 e/\AA$^3 $ for (b) and (c).
Insets schematically show the corresponding atomic orbitals of Cr and As.}
\end {figure}

In the spin-down channel, Fe:3d majority states lie deeply in
the valence band and are fully occupied, as shown in Figure 2c.
Most of the minority states of Cr are above $E_{\textrm F}$ and form the
bottom of conduction band, in contrast with
Fe in the spin-up channel, due to the difference in the electronegativity
between Cr and Fe. Around the top of valence band
there are noticeable contributions from Cr:3d minority states as indicated
in Figure 2b. It is intriguing to notice that the
lattice structure of the present iron-pnictide material is crucial for
opening a gap of $\sim 0.3$ eV between Cr:3d states in the spin-down channel.
Since the Cr:3d$_{xy}$ orbitals reside on the TM
plane with their petals pointing to the nearest neighboring Cr ions
along (110) or (1-10) direction, and the As atoms just locate
above the center of the TM squares (see inset of Figure 3a), the As:4p$_{z}$
orbitals mediate the exchange interaction between the
Cr:3d$_{xy}$ states. Meanwhile, the As:4p$_{y}$ (As:4p$_{x}$) state
provides an evident bridge for the exchange coupling between the two
Cr:3d$_{xz}$ (Cr:3d$_{yz}$) states.
As the result, the spin-down As:4p states in the vicinity of $E_{\textrm F}$
hybridize with a part of the Cr:t$_{2g}$ minority states (Figure 2b
and 2d), and lower their energies to the valence band. These features are well captured
by the band-decomposed charge density \cite{vasp-cite-1,vasp-cite-2}
depicted in Figure 3 obtained by calculations based on
the $\sqrt{2}\times\sqrt{2}\times1$ cell.

\begin{figure*}[t]
\includegraphics[width=14cm]{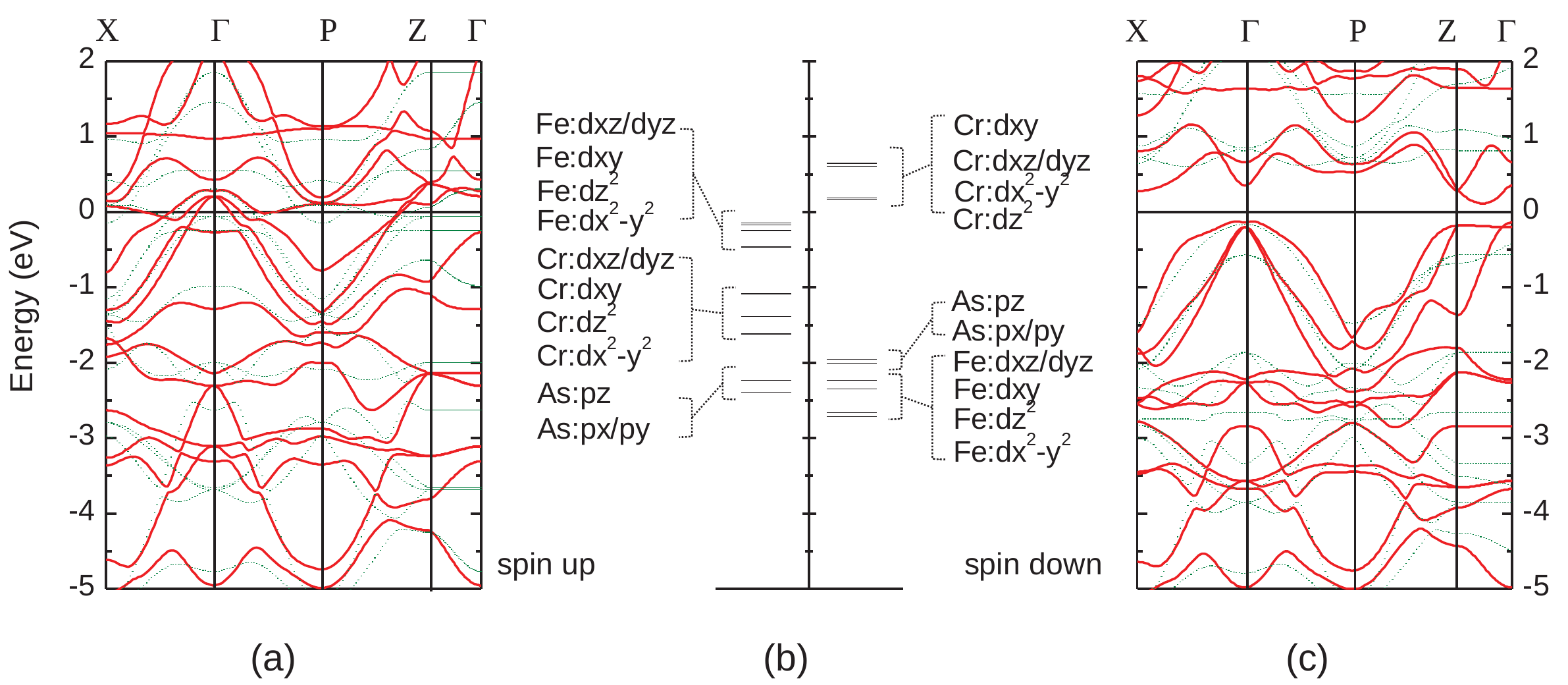}
\label{bands}
\caption{Band structures of
BaCrFeAs$_{2}$ with the red solid and green dotted curves denote
results based on the first-principles calculations and the
TB model, respectively. (a)/(c) for spin-up/down channel;
(b) for on-site energies of relevant orbitals of the TB
model derived from the MLWF calculation. }
\end{figure*}

\begin{table}
\label{tab:table1}
\caption{ Hopping integrals between the orbitals of nearest-neighboring
As-As, As-Cr/Fe and Cr-Fe ions derived from the MLWF calculation in the units of eV \textsuperscript{\emph{a}}}
\begin{tabular}{ccccccc}
\hline
   orbitals  & \multicolumn{2}{c}{hopping integrals}  \\
             &      Spin-up     &     Spin-down       \\
\hline
As:p$_{x}$/p$_{y}$-Cr:d$_{yz}$/d$_{xz}$             &0.70 &0.80 \\ 
As:p$_{z}$-Cr:d$_{xy}$                              &0.61 &0.68 \\ 
As:p$_{x}$/p$_{y}$-Cr:d$_{x^{2}-y^{2}}$             &0.50 &0.65 \\ 
As:p$_{z}$-Cr:d$_{3z^{2}-r^{2}}$                    &0.50 &0.65 \\ 
As:p$_{x}$/p$_{y}$-Cr/Fe:d$_{3z^{2}-r^{2}}$         &0.24 &0.27 \\ 
As:p$_{x}$/p$_{y}$-Fe:d$_{yz}$/d$_{xz}$             &0.66 &0.58 \\ 
As:p$_{z}$-Fe:d$_{xy}$                              &0.57 &0.50 \\ 
As:p$_{x}$/p$_{y}$-Fe:d$_{x^{2}-y^{2}}$             &0.50 &0.39 \\ 

As:p$_{x}$/p$_{y}$-As:p$_{x}$/p$_{y}$ (\uppercase\expandafter{\romannumeral1})   &0.27 &0.27 \\ 
As:p$_{x}$-As:p$_{y}$ (\uppercase\expandafter{\romannumeral1})                   &0.34 &0.33 \\ 
As:p$_{z}$-As:p$_{z}$ (\uppercase\expandafter{\romannumeral2})                   &0.29 &0.28 \\ 
As:p$_{z}$-As:p$_{x}$/p$_{y}$ (\uppercase\expandafter{\romannumeral2})           &0.41 &0.38 \\ 
As:p$_{x}$/p$_{y}$-As:p$_{x}$/p$_{y}$ (\uppercase\expandafter{\romannumeral2})   &0.34 &0.30 \\ 
As:p$_{z}$-As:p$_{z}$ (\uppercase\expandafter{\romannumeral3})                   &0.73 &0.76 \\ 

Cr/Fe:d$_{x^{2}-y^{2}}$-Fe/Cr:d$_{3z^{2}-r^{2}}$    &0.22 &0.21 \\ 
Cr:d$_{xz}$/d$_{yz}$-Fe:d$_{xz}$/d$_{yz}$           &0.28 &0.28 \\ 
Cr:d$_{x^{2}-y^{2}}$-Fe:d$_{x^{2}-y^{2}}$           &0.37 &0.37 \\ 
Cr:d$_{xy}$-Fe:d$_{xy}$                             &0.27 &0.26 \\ 
\hline
\end{tabular}
\\
\textsuperscript{\emph{a}} The Roman numbers denote the hoppings between As-As ions (\uppercase\expandafter{\romannumeral1})
within the same As plane, (\uppercase\expandafter{\romannumeral2}) crossing the TM plane
and (\uppercase\expandafter{\romannumeral3}) along the c axis, which can be read from Figure 1.
\end{table}

\begin{table}
\label{tab:table2}
\caption{Onsite energy of As:4p and TM:3d orbitals for the TB
model derived from the MLWF calculation in the units of eV}
\begin{tabular}{ccccc}
            & Cr:d$_{3z^{2}-r^{2}}$ & Cr:d$_{xz}$/d$_{yz}$ & Cr:d$_{x^{2}-y^{2}}$ & Cr:d$_{xy}$ \\
  \hline
  Spin up   & -1.61 & -1.08 & -1.62 & -1.38  \\
  Spin down &  0.18 &  0.61 &  0.20 &  0.65  \\
\end{tabular}
\\
\begin{tabular}{ccccc}
            & Fe:d$_{3z^{2}-r^{2}}$ & Fe:d$_{xz}$/d$_{yz}$ & Fe:d$_{x^{2}-y^{2}}$ & Fe:d$_{xy}$ \\
  \hline
  Spin up   & -0.24 & -0.17 & -0.46 & -0.14 \\
  Spin down & -2.66 & -1.96 & -2.71 & -2.00 \\
\end{tabular}
\\
\begin{tabular}{ccc}
            & As:p$_{z}$ & As:p$_{x}$/p$_{y}$ \\
  \hline
  Spin up   & -2.24 & -2.39 \\
  Spin down & -2.23 & -2.34 \\
\end{tabular}
\end{table}

\subsection{Tight-binding Model}

For a better understanding on the novel HMAFM property of BaCrFeAs$_{2}$,
in which many orbitals contribute around $E_{\textrm F}$,
it is helpful to build a TB picture. For
this purpose, we adopt the MLWF method \cite{PWscf,Wannier90}, which yields both
on-site energies and hopping integrals of the TB model.
Since the whole TM:3d and As:4p orbitals are involved in the
covalency, all of them are treated as the initial guess for orbital
projections. It is confirmed that the DFT results can be perfectly
reproduced. We then reduce the hopping
ranges to the minimum with which one can still capture the main
features of the DFT results. In this way, the important orbitals
which govern the band dispersion around $E_{\textrm F}$ are
identified as TM:t$_{2g}$ and As:4p, and the hopping integrals are for
the nearest neighbors only.

The numerical results for the hopping
integrals and the on-site energies are given in Table I and II respectively,
and the band structures derived from the TB model are shown in
Figure 4 in fairly good agreement with the VASP results.
The onsite energies of spin-up Fe:3d states are slightly below
$E_{\textrm F}$, while the spin-down Cr:3d orbitals are slightly higher than
$E_{\textrm F}$.  It is clear that this difference is responsible for the
half metallic feature of the present material BaCrFeAs$_{2}$.

As listed in Table I, the As:p$_{z}$-As:p$_{z}$ hopping along c axis is quite large
($\sim$ 0.7 eV) due to the crystal structure of the present '122' system
with the shortest As-As bond along the c axis. This interaction is crucial
for the c axis AFM order of BaCrFeAs$_{2}$, and thus the feature of HMAFM,
in contrast to the case of the so-called '1111' system \cite{nakao-1111}.
Therefore we predict that the present BaCrFeAs$_{2}$ as well
as other '122' materials are the most promising candidates for
HMAFM among the iron-pnictides.

\section{Conclusions}

To summarize, based on first-principles calculations and a TB
model we predict that
BaCrFeAs$_{2}$ is a candidate material for half-metallic antiferromagnet.
The novel half-metallic band structure of this material is generated
by the selective, strong hybridization between As:4p states and minority
Cr:3d states, due to the difference in the electronegativity between Fe and Cr.
Both the density-functional calculations and the TB
analysis reveal clearly that the crystal structure of iron-pnictide
is crucial for opening a gap around $E_{\textrm F}$ as large as
$\sim$ 0.3 eV in the single spin channel. The zero
magnetization is achieved by the magnetic compensation of Cr, Fe and
As atoms. It is highly anticipated that the theoretical
prediction formulated in the present work can be checked experimentally.
The endeavor will then open a new field of iron-based spintronics.

\begin{acknowledgments}
The authors thank I. V. Solovyev, B. Liu and Y.-M. Nie for discussions.
The calculations were performed on Numerical Materials Simulator (SGI Altix) of NIMS.
This work was supported by WPI Initiative on Materials Nanoarchitronics, MEXT, Japan.
\end{acknowledgments}

%

\end{document}